\documentclass[preprint,showpacs,preprintnumbers,amsmath,amssymb]{revtex4}

\usepackage{graphicx}
\usepackage{dcolumn}
\usepackage{bm}

\begin{document}


\title{Anderson Localization in Quark-Gluon Plasma}

\author{Tam\'as G.\ Kov\'acs\footnote{Supported by OTKA Hungarian Science
Fund grants 46925 and 49652 and EU Grant (FP7/2007-2013)/ERC n$^o$208740. I
also thank S.D.\ Katz and D.\ N\'ogr\'adi for stimulating discussions.} and
Ferenc Pittler}
\affiliation{%
Department of Physics, University of P\'ecs \\
H-7624 P\'ecs Ifj\'us\'ag u.\ 6, Hungary 
}%


\date{\today}

\begin{abstract}

At low temperature the low end of the QCD Dirac spectrum is well described by
chiral random matrix theory. In contrast, at high temperature there is no
similar statistical description of the spectrum. We show that at high
temperature the lowest part of the spectrum consists of a band of
statistically uncorrelated eigenvalues obeying essentially Poisson statistics
and the corresponding eigenvectors are extremely localized. Going up in the
spectrum the spectral density rapidly increases and the eigenvectors become
more and more delocalized. At the same time the spectral statistics gradually
crosses over to the bulk statistics expected from the corresponding random
matrix ensemble. This phenomenon is reminiscent of Anderson localization in
disordered conductors. Our findings are based on staggered Dirac spectra in
quenched SU(2) lattice simulations.

\end{abstract}

\pacs{11.15.Ha,12.38.Gc,12.38.Aw,11.30.Rd}
\maketitle


The spectrum of the QCD Dirac operator contains important information
regarding the properties of strongly interacting physical systems. Statistical
properties of the spectrum completely determine the bulk thermodynamical
observables. The low end of the Dirac spectrum is particularly important since
that part dominates the quark propagators. The statistics of low Dirac
eigenvalues is fundamentally different in the low temperature chirally broken
phase and the high temperature chirally symmetric phase.  According to the
Banks-Casher formula \cite{Banks:1979yr} the spectral density around zero is
proportional to the order parameter of chiral symmetry breaking. Therefore in
the chirally symmetric phase it vanishes whereas in the broken phase it is
non-zero. In the latter case random matrix theory (RMT) provides an
essentially complete statistical description of the low lying part of the
Dirac spectrum \cite{Verbaarschot:2000dy}. In the intermediate volume, so
called epsilon regime the RMT behavior is well understood analytically
through an effective sigma model description and numerically through lattice
simulations.

In contrast, above the finite temperature transition, $T_c$, in the
chirally symmetric phase there is no well understood statistical description
of the low eigenvalues of the Dirac operator. In this regime, lacking any
analytical insight, one can regard the Dirac operator as a randomly
fluctuating matrix of size going to infinity in the thermodynamic limit. From
this perspective there are two fundamentally different possibilities for the
spectrum of the Dirac operator. If typical random fluctuations can freely mix
eigenvectors, eigenvectors become extended and the eigenvalue statistics is
described by the corresponding RMT. If on the other hand, fluctuations in the
matrix elements cannot mix the eigenvectors, in some basis they become
localized and the eigenvalues become independent, obeying essentially Poisson
statistics. Lattice simulations can test which scenario happens in reality.

Above $T_c$ the spectral density vanishes at zero and RMT has
predictions for the eigenvalue statistics at such a ``soft edge''
\cite{Forrester}. Lattice simulations, however, did not find agreement with
these predictions \cite{Farchioni:1999ws,Damgaard:2000cx}. Another earlier
study \cite{Pullirsch:1998ke} did not focus on the spectrum edge, but
considered full Dirac spectra and found bulk RMT statistics
for the full spectrum.  More recently the possibility of Poisson eigenvalue
statistics was suggested again in Ref.\ \cite{GarciaGarcia:2006gr}. Based on
lattice simulations the authors argued that around $T_c$ the low temperature
RMT statistics is gradually deformed towards Poisson statistics.
Ref.\ \cite{Gavai:2008xe}, based again on lattice simulations, found that
although low lying Dirac eigenvectors become very localized above $T_c$, the
localization is not stable and most likely will disappear in the thermodynamic
limit. This finding would disfavor the appearance of Poisson statistics in the
spectrum. Very recently, using overlap fermion lattice simulations,
Ref.\ \cite{Kovacs:2009zj} found Poisson behavior for the lowest two
eigenvalues.

In the present paper we offer a complete understanding of this rather unclear
situation. We show that above $T_c$ the lowest part of the spectrum consists of
statistically independent eigenvalues obeying Poisson statistics and the
corresponding eigenvectors are extremely localized. Going up in the spectrum
the spectral density rapidly increases and the eigenvectors get
delocalized. At the same time the spectral statistics gradually crosses over
to the bulk statistics expected in the corresponding random matrix
ensemble. We also show that the number of Poisson type eigenvalues depends
only on the physical temperature and the physical spatial volume 
and not on the lattice spacing.

The phenomenon we report here is analogous to Anderson localization
occurring in crystalline conductors in the presence of disorder. In
that case disorder causes the appearance of localized electron states at
the band edge, but for sufficiently weak disorder states towards the band
center remain delocalized \cite{Lee:1985zzc}. The corresponding eigenvalue
statistics changes from Poisson around the band edge to random matrix
statistics towards the band center \cite{Evers:2008zz}. It is interesting to
note that while in disordered conductors localization is most often
understood in terms of disorder in the diagonal (on-site) matrix elements of the
Hamiltonian, in the case of QCD, disorder is entirely in the hopping terms. In
fact the on-site matrix elements of the staggered Dirac operator are
identically zero and all ``disorder'' is in the gauge field in the
hopping terms.

The low lying QCD Dirac spectrum is known to depend strongly on the temporal
fermionic boundary condition which is effectively a combination of the
Polyakov loop and the explicitly chosen anti-periodic boundary condition
\cite{Bilgici:2009tx}.  In the quenched $SU(2)$ theory the Polyakov loop
$Z(2)$ symmetry is spontaneously broken above $T_c$. Although in the quenched
theory the two sectors are equivalent, here we only use configurations in the
``physical'' Polyakov loop sector, the one that would survive in the presence
of dynamical fermions. We use fermion boundary conditions that are
anti-periodic in the time direction and periodic in all the spatial
directions.

At first we summarize the details of the numerical simulations. The data is
based on quenched simulations of the $SU(2)$ gauge theory with Wilson
plaquette coupling $\beta=2.6$ and time extension $N_t=4$. This corresponds to
a temperature of $T=2.6T_c$, well above the finite temperature phase
transition.  The simulations were done at four different spatial volumes,
$N_s^3=16^3,24^3,32^3,48^3$. To assess what happens in the continuum limit we
included an additional $N_t=6, N_s=36, \beta=2.725$ ensemble matched
to the $N_s=24$ simulation in terms of physical box size and
temperature, but on a 1.5 times finer lattice. On these ensembles we computed
the 256 smallest positive eigenvalues of the staggered Dirac operator. Due to
the exact twofold degeneracy of the eigenvalues this yields 128 independent
positive eigenvalues per configuration that we used for the the statistical
analysis.
 
\begin{figure}
\includegraphics[width=0.8\columnwidth,keepaspectratio]{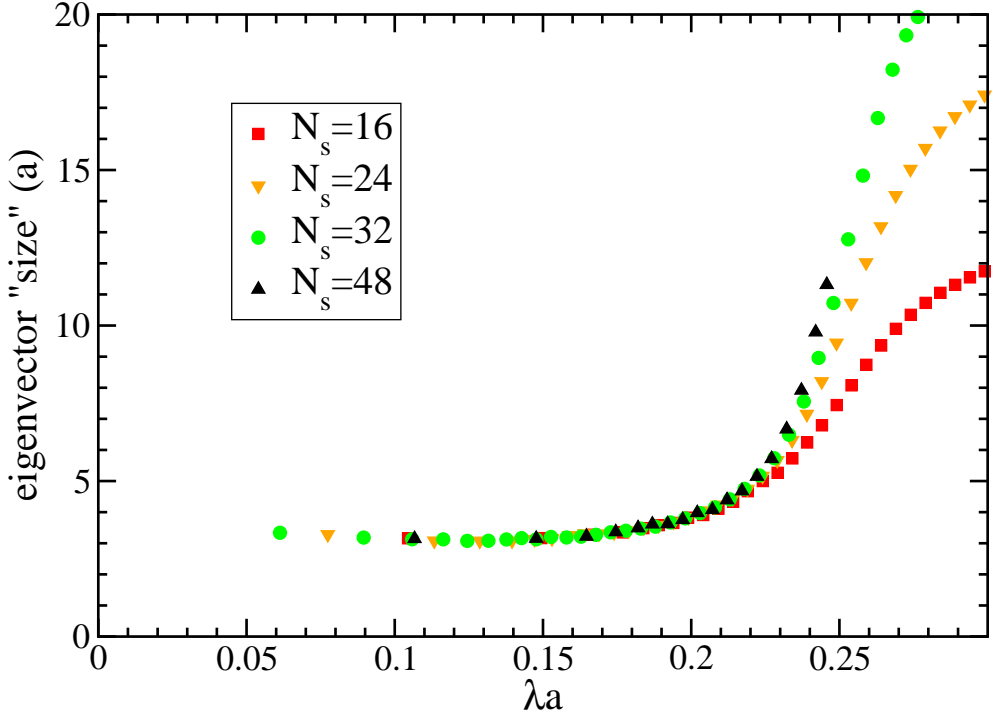}
\caption{\label{fig:em_size} The average linear extension of
  eigenvectors as a function of the corresponding eigenvalues in different
  spectral windows. The linear size is based on the participation ratio. The
  different curves correspond to different spatial box sizes
  $N_s=16,24,32,48$.}
\end{figure}

As a first step we studied the spatial localization of these eigenmodes. A
possible way to measure that is through the participation ratio
\begin{equation}
 {\cal V} = \left[ \sum_x (\psi^\dagger \psi(x))^2 \right]^{-1},
\end{equation}
where $\psi$ is the normalized eigenvector. If the eigenvector spreads
uniformly in a four-volume ${\cal V}$ and is zero elsewhere then its
participation ratio is ${\cal V}$. Assuming that at this high temperature the
low eigenvectors can maximally spread in the time direction one can define a
length scale
\begin{equation}
 d = \left[ \frac{\cal V}{N_t} \right]^{1/3}
\end{equation} 
characterizing the spatial extension of the eigenvectors. In
Fig.\ \ref{fig:em_size} we plot how this quantity changes in the spectrum. The
averages were calculated in non-overlapping spectral windows separately for
different spatial box sizes. It is apparent that the lowest eigenmodes up to
about $\lambda a<0.22$ are very localized and their spatial extension is
independent of the box size. Above that point the
eigenvectors rapidly start to delocalize and their spatial size
becomes dependent on the box size. This is the point in the spectrum
that is known as the mobility edge in the context of Anderson localization.  
     
\begin{figure}
\vspace{1cm}
\includegraphics[width=0.8\columnwidth,keepaspectratio]{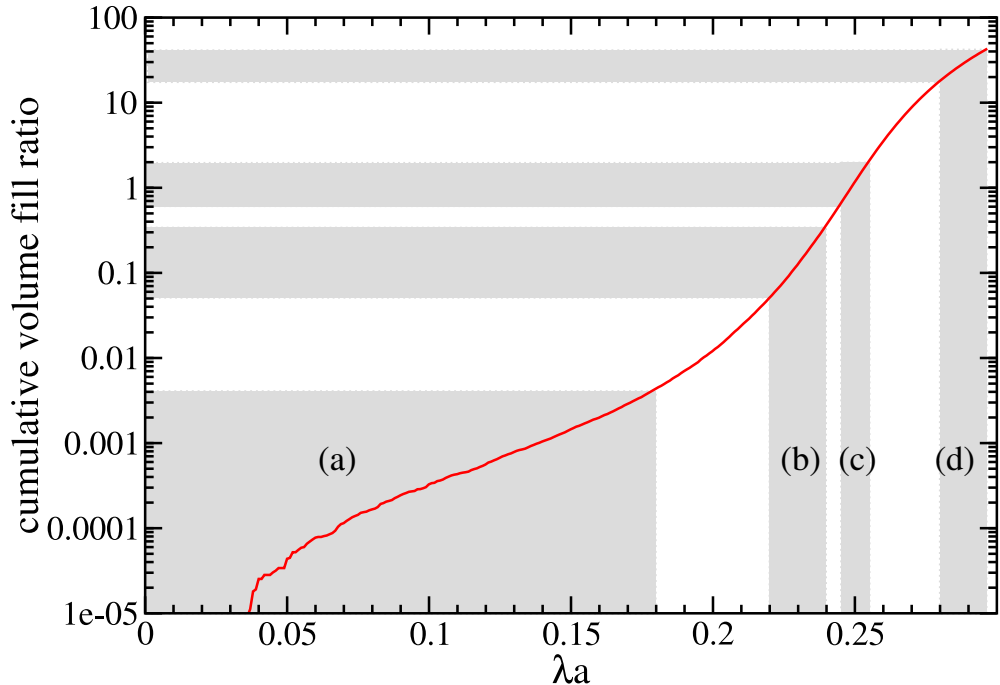}
\caption{\label{fig:evsvolume24} The cumulative volume fill fraction for the
  $24^3\times 4$ ensemble. The unfolded level spacing distribution will be
  computed  separately in the four shaded spectral windows marked by (a)-(d)
  (see Fig.\ \ref{fig:ks_levsp}).} 
\end{figure}

Besides the localization of eigenvectors the other important factor determining
how easily eigenvectors can mix is the spectral density. A useful
quantity that reflects the combined effect of localization and
spectral density can be defined as follows. For each eigenvector the
participation ratio is an approximate measure of the four-volume
occupied by the given eigenvector. We call the cumulative volume fill fraction
the sum of the volumes occupied by all the eigenvectors corresponding to
eigenvalues less than $\lambda$. It is understood to be normalized by the
total box four-volume. In Fig.\ \ref{fig:evsvolume24} we show the volume fill
fraction as a function of the eigenvalue.
It is a rapidly increasing function of $\lambda$. In the lowest part of the
spectrum where the spectral density is small and eigenvectors are localized
the fill fraction is much smaller than unity. These eigenvectors have very
little spatial overlap and they are essentially produced independently in
different sub-volumes. The corresponding eigenvalues are expected to be
independently distributed and obey Poisson statistics.  In contrast, 
above $\lambda a>0.28$ the volume fill fraction is much
bigger than unity and the eigenvectors here strongly overlap. Therefore they
can freely mix and from this point up in the spectrum the eigenvalue
statistics is expected to be described by random matrix theory. Between the
two extremes there must be some transition from Poisson to random matrix
statistics.

A simple way of testing these expectations is to consider the unfolded level
spacing distribution computed in spectral windows located in the three above
described regimes. Unfolding is a simple mapping of the eigenvalues that
eliminates all the information about the spectral density which is not
universal, but keeps universal eigenvalue correlations intact. Numerically we
unfolded by ordering all the eigenvalues by magnitude on all
configurations in the given ensemble and mapping each eigenvalue to its rank
order normalized by the number of configurations in the given ensemble. By
construction the eigenvalues transformed in this way have constant unit
spectral density. 

If the original eigenvalues are uncorrelated, the unfolded level spacing
distribution is expected to be a simple exponential $P(x)=e^{-x}$. If, on the
other hand, the original eigenvalues obey random matrix statistics the
unfolded level spacing distribution should follow the so called Wigner
surmise of the corresponding random matrix ensemble
\cite{Verbaarschot:2000dy}.  The staggered Dirac operator with fermions in the
$SU(2)$ fundamental representation correspond to the chiral symplectic random
matrix ensemble and the Wigner surmise in that case is \cite{Halasz:1995vd}

\begin{equation}
 P(x) = \frac{2^{18}}{3^6\pi^3} \, x^4 \, 
               \exp\left( -\frac{64}{9\pi} x^2 \right).
\end{equation}
\begin{figure}
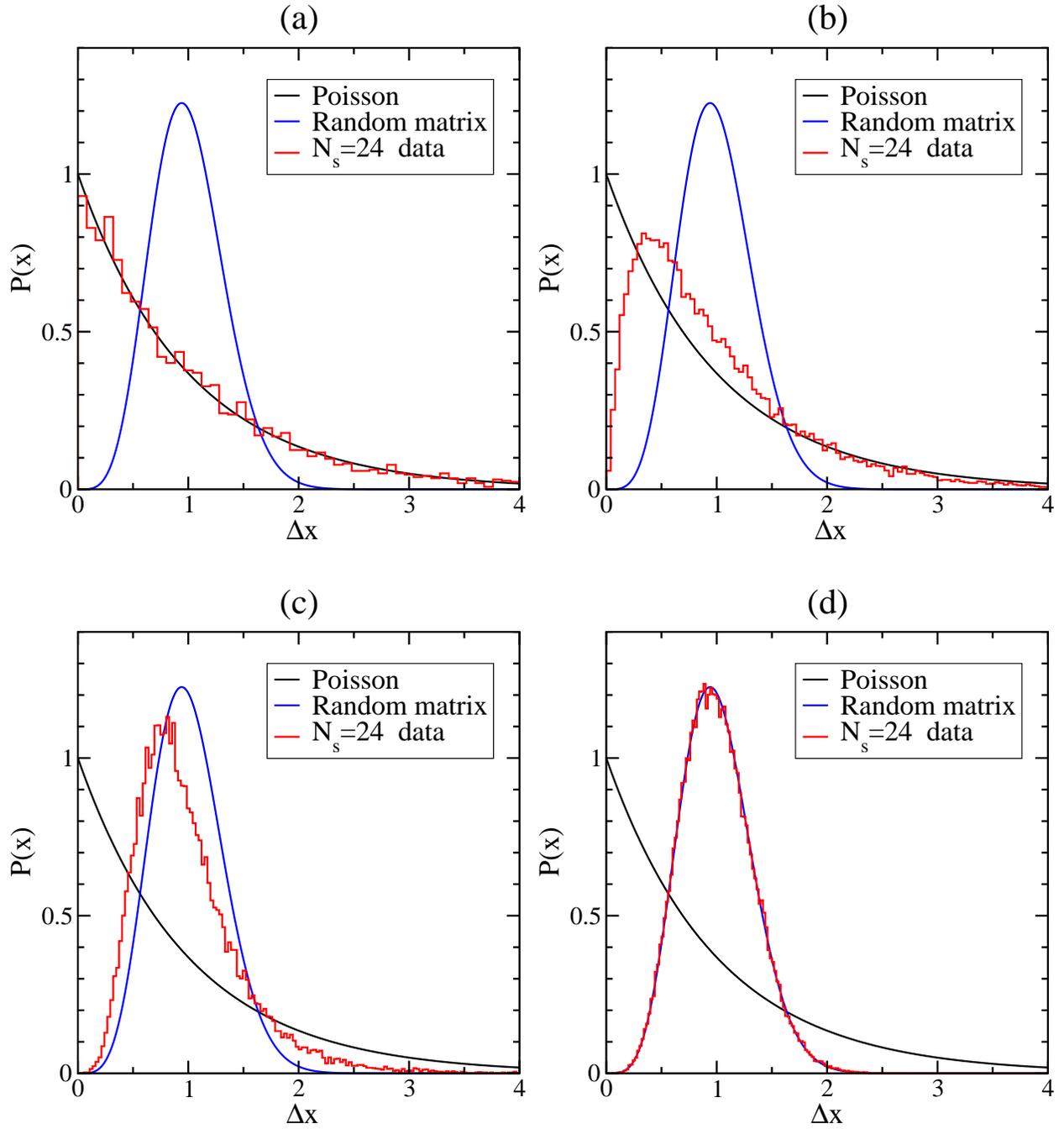

\vspace{1cm}
\begin{tabular}{ll}
\includegraphics[width=0.50\columnwidth,
                       keepaspectratio]{Figs/ks_levsp_0.00-0.18.eps} &
\includegraphics[width=0.50\columnwidth,
                       keepaspectratio]{Figs/ks_levsp_0.22-0.24.eps} \\[5mm]
\includegraphics[width=0.50\columnwidth,
                       keepaspectratio]{Figs/ks_levsp_0.245-0.255.eps} &
\includegraphics[width=0.50\columnwidth,
                       keepaspectratio]{Figs/ks_levsp_0.28-0.297.eps}
\end{tabular}
\caption{\label{fig:ks_levsp} The panels show the unfolded level spacing
  distribution in different regions of the spectrum. The labeling (a)-(d)
  corresponds to the spectral windows indicated in Fig.\ \ref{fig:evsvolume24}
  with the shaded areas. The curved lines are the exponential
  distribution and the Wigner surmise.}
\end{figure}
In Fig.\ \ref{fig:ks_levsp} we plot the unfolded level spacing distribution
averaged separately for the four spectral windows indicated in
Fig.\ \ref{fig:evsvolume24}. For comparison we also show the exponential and
Wigner surmise distributions expected if the level statistics is Poisson and
random matrix respectively. Going upwards in the spectrum the transition from
Poisson to random matrix statistics is obvious.  

We note that the authors of Ref.\ \cite{Damgaard:2000cx} also computed the
unfolded level spacing statistics between the first and the second, the second
and the third, etc.\ eigenvalues separately and found that for the lowest few
eigenvalues there was a slight tilting of the distribution to the left
compared to the Wigner surmise (like the one seen in our
Fig.\ \ref{fig:ks_levsp}c). However, they could not see the Poisson statistics
we report here because they used much smaller spatial volumes than what we are
using here. Since the spectral density in the Poisson regime is small, in a
small volume even the first few eigenvalues are in the random matrix regime
most of the time.

So far we discussed level spacing statistics for a given spatial box size
$(N_s=24)$ and a given coarseness of the lattice. For these results to
represent real physics it is important to check what happens in the
thermodynamic and in the continuum limit. In the remainder of the paper we
address these two questions.

\begin{figure}
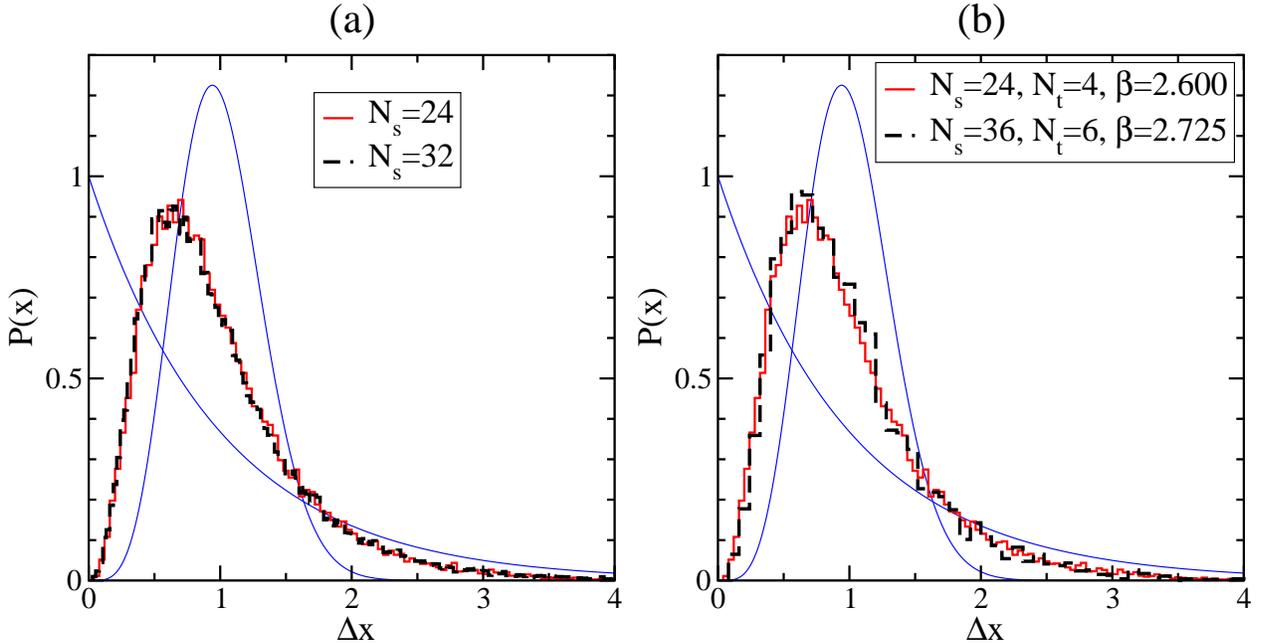

\vspace{1cm}
\begin{tabular}{ll}
\includegraphics[width=0.5\columnwidth,keepaspectratio]{Figs/levsp_volscale.eps}
 & 
\includegraphics[width=0.5\columnwidth,keepaspectratio]{Figs/levsp_cont.eps}
\end{tabular}
\caption{\label{fig:levsp_scale} The unfolded level spacing distribution for
  two different spatial box sizes at fixed $N_t$ (a) and for two different
  $N_t$'s at fixed physical temperature and physical box size (b). }
\end{figure}

{\em Thermodynamic limit---}We have already seen that the lowest eigenmodes
are very localized and their spatial extension is not affected by the volume
of the box. Therefore we expect that these modes occur independently and their
average number is proportional to the spatial volume. More generally
eigenvalues following any intermediate statistics between Poisson and random
matrix are also expected to occur in numbers proportional to the spatial
volume. If this is true the statistics of eigenmodes number $n_1$ through
$m_1$ in a spatial volume $V_1$ should be the same as that of eigenvalues
$n_2$ through $m_2$ in a spatial volume $V_2$ provided that
$\frac{n_1}{n_2}=\frac{m_1}{m_2}=\frac{V_1}{V_2}$.  We verified this by
comparing the unfolded level spacing distribution in two different spatial box
sizes, $N_s=24$ and $32$.  In Fig.\ \ref{fig:levsp_scale}a we compare the
unfolded level spacing distribution of eigenvalues 10-20 in the smaller volume
and 24-47 in the bigger volume and find good agreement.  To demonstrate that
these eigenvalues fall in the intermediate statistics regime we also indicated
in the Figure the exponential and Wigner surmise distributions with thin solid
lines. This shows that the density of eigenvalues with a given intermediate
statistics scales with the spatial volume as expected.

{\em Continuum limit---}To see how this picture changes towards the continuum
limit we considered an additional ensemble on a finer lattice. The parameters
of that, $N_t=6,\, N_s=36,\, \beta=2.725$ were tuned to match the physical
temperature and spatial volume of the $24^3\times 4$ ensemble. Since the
average smallest eigenvalue changes roughly with the smallest Matsubara
frequency, it does not make sense to match the spectral windows of these
ensembles. Instead we computed the unfolded level spacing distribution of
eigenvalues 10-20 on each configuration for both ensembles. As can be seen in
Fig.\ \ref{fig:levsp_scale}b there is perfect agreement between the two
ensembles.

On the one hand, this means that the deformation of the distribution from
Poisson to random matrix statistics occurs through a universal path
independently of the lattice spacing. On the other hand, it also implies that
the number of very localized Poisson eigenmodes is proportional to the
physical spatial volume and not the volume in lattice units. It could suggest
that these eigenmodes are localized on some physical gauge field objects that
survive the continuum limit with a finite physical density. This scenario is
also supported by the fact that the localization range of the eigenmodes in
physical units, computed from the participation ratio was roughly the same on
the coarser and finer ensemble.

{\em Discussion---}Mixing instanton anti-instanton zero modes could be a
natural candidate to explain the localized modes. Using the number of
zero modes of the overlap Dirac operator on these ensembles we estimated the
density of topological objects. We found that their density is more than an
order of magnitude smaller than what would be needed to explain the Poisson
modes. No matter what their origin is, however, it is clear that the fermion
boundary condition plays a crucial role in their appearance because in the
opposite $SU(2)$ Polyakov loop sector these modes are completely absent
\cite{Kovacs:2008sc}. We hope to return to a more detailed discussion of the
role of the boundary condition in a later publication.

Another important issue is how universal the appearance of Poisson modes
is. In Ref.\ \cite{Kovacs:2009zj} in a range of spatial volumes the lowest two
eigenvalues of the overlap Dirac operator were seen to be Poissonian.
In a much smaller statistics study than reported here with
staggered fermions we also verified that the overlap spectrum has
similar behavior of crossing from Poisson to orthogonal random matrix
statistics, the RMT corresponding to the symmetries of the overlap operator. It
means that the phenomenon we found does not rely on a particular lattice
fermion formulation.

\end{document}